\documentclass[aps,preprint,floatfix,superscriptaddress,amsmath,showpacs,amssymb]{revtex4}

\usepackage{graphicx}
\usepackage{epsfig}

\begin{document}

\title{Excitation levels and magic numbers of small para-Hydrogen clusters (N$ \le 40$)}
\date{\today}

\author{Rafael Guardiola}
\affiliation{ Departamento de F\'{\i}sica At\'omica y Nuclear, Facultad de
F\'{\i}sica, 46100 Burjassot, Spain}

\author{Jes\'us Navarro}
\affiliation{ IFIC (CSIC-Universidad de Valencia), Apdo. 22085,
46071 Valencia, Spain}


\begin{abstract}
The excitation energies of parahydrogen clusters have been systematically
calculated by the diffusion Monte Carlo technique in steps of one molecule
from 3 to 40 molecules. These clusters possess a very rich spectra, with
angular momentum excitations arriving up to $L=13$ for the heavier ones.
No regular pattern can be guessed in terms of the angular momenta and the
size of the cluster. Clusters with $N=13$ and 36 are characterized by a
peak in the chemical potential and a large energy gap of the first excited
level, which indicate the magical character of these clusters. 
From the calculated excitation energies the partition function has been obtained, 
thus allowing for an estimate of thermal effects. An enhanced production is 
predicted for cluster sizes $N=13$, 31 and 36, in agreement with experiment. 
\end{abstract}

\maketitle

\section{Introduction}

Small (pH$_2$)$_N$ clusters of parahydrogen have been produced in a cryogenic
free jet expansion and studied by Raman spectroscopy~\cite{tejeda2004}.
The $Q(0)$ Raman line of the H$_2$ monomer is shifted as the number $N$ of
molecules in the cluster changes, thus providing a method to identify the
cluster mass. The first seven resolved peaks next the monomer line have been
assigned to clusters with $N=2,\dots,8$ molecules. Although in that experiment
the resolution was not enough to resolve larger sizes, broad maxima were 
observed at $N\approx13$ and 33, and perhaps 55, which have been interpreted
as a propensity for geometric shell structures. Indeed, classical 
static~\cite{hoare1971,hoare1979,honeycutt1987} and molecular 
dynamics~\cite{honeycutt1987,jellinek1986} results, based on a generic
Lennard-Jones interaction potential, indicate that the expected structures
for such clusters are the so-called Mackay icosahedra~\cite{mackay1962},
exhibiting some magic sizes (13, 33, 55...) related to the packing of
molecules in closed icosahedral arrangements. A very complete discussion
of classical geometrical patterns and their relation with the interaction
features can be found in Ref.~\cite{baletto2005} and references herein.
However, quantum effects play a major role in pH$_2$ clusters,
like in their analogous helium droplets.
The path integral Monte Carlo (PIMC) simulations of Sindzingre
{\em et al.}~\cite{sindzingre1991} have shown indeed that the superfluid
fraction in pH$_2$ clusters with 13 and 18 molecules become large at
temperatures below $T \simeq 2$~K. This prediction prompted both
experimental~\cite{grebenev2000} and
theoretical~\cite{kwon2002,paesani2003,tang2004,paesani2005} research
of small clusters consisting of pH$_2$ molecules surrounding an OCS
chromophore, confirming the existence of a superfluid response.

Quantum Monte Carlo (QMC) methods have been widely used in recent years
as a theoretical tool to study pH$_2$ clusters. Several QMC techniques
have been employed to calculate their properties, namely
variational Monte Carlo (VMC) and diffusion Monte Carlo 
(DMC)~\cite{rama1991,mcmahon1993,mcmahon1994,guardiola2006a,guardiola2008},
path integral Monte Carlo
(PIMC)~\cite{sindzingre1991,scharf1992,mezzacapo2006,mezzacapo2007,khairallah2007},
reptation Monte Carlo (RMC)~\cite{baroni2005} and path integral ground
state (PIGS) Monte Carlo~\cite{cuervo2006}.
Only some specific values of the number of constituents $N$ have been
considered in the past, related to the expected MacKay icosahedra structures.
More recently, systematic calculations as a function of the number of
molecules in the cluster have been performed. The ground state energies
and the one-body densities of pH$_N$ clusters have been calculated by the
DMC technique in steps of one molecule from N=3 to 50~\cite{guardiola2006a},
by the PIGS technique from N=2 to 20~\cite{cuervo2006} and by the PIMC from
N=5 to 40~\cite{khairallah2007,mezzacapo2007}. The calculations show that
(pH$_2$)$_N$ clusters exhibit a clear geometrical order, with the
molecules occupying concentric spherical shells, which could be
related to some polyhedric arrangement. The apparent incompatibility between
the large superfluid fractions and the structured radial distribution
densities has been recently clarified~\cite{khairallah2007,mezzacapo2007}
by PIMC calculations which show that superfluidity is localized at the
surface of the clusters.

Whereas up to $N\simeq22$ these calculations are substantially in agreement,
for heavier clusters there are noticeable differences between DMC and PIMC
results, particularly for $N \geq 26$. PIMC chemical potentials show very
prominent peaks at N=26, 29, 34 and 39, in contrast with a smoother
behavior obtained with DMC. These differences subsist even after an improved
DMC calculation~\cite{guardiola2008}, and seem to be related to
thermal effects. According to Mezzacapo and Boninsegni~\cite{mezzacapo2007} they
should be associated to a coexistence of solid-like and liquid-like phases,
with a dominance of the latter at low T, as a result of both the zero-point
motion and quantum permutation exchanges.
However, thermal effects could manifest in enhanced stability thresholds
at finite temperature, similarly to what has been observed in
$^4$He droplets~\cite{bruhl2004}. Such droplets are definitely
liquid-like and nevertheless they exhibit some magic numbers, which
are not related to enhanced ground state binding energies at specific
values of $N$. They are instead stability
thresholds, related to the cluster sizes at which excited levels cross the
chemical potential curve and become stabilized~\cite{guardiola2006}.
An analysis of the possible existence of similar enhanced stabilities in
pH$_2$ could be helpful to interpret the experimental results of
Ref.~\cite{tejeda2004}

The purpose of the present article is to provide an accurate microscopic
description of the ground state and the low-lying excited levels of
(pH$_2$)$_N$ clusters. In the past, the excited states of the $N=7$ cluster
were analyzed by McMahon {\it et al.}~\cite{McMahon1993}. We present here
a systematic calculation of excitations with angular momentum from $L=0$
to 13 for clusters with $N=3,\dots,40$. For a given $L$-state the
Schr\"odinger equation has been solved by means of stochastic DMC techniques using
the pairwise potential determined by Buck {\it et al.}\cite{buck1983},
hereafter referred to as BHKOS. Another popular choice is the potential
due to Silvera and Goldman~\cite{silvera1978}, hereafter referred to as SG.
Both potentials combine {\em ab initio} calculations with properties of the
gas (or solid) as well as experimental information from collisions.
The main difference among them is that SG contains a repulsive long-range
term ($c_9/r^9$) to approximate the effective potential in a solid.
As showed in Ref.~\cite{guardiola2008}, the BHOKS potential provides more
binding than the SG one, and the BHOKS chemical potential is slightly
higher than the SG one.

The paper is organized as follows. In Sec. II, technical details of the
DMC calculations are given for the ground state and low-lying excited states.
The energy levels of (pH$_2$)$_N$ clusters are presented and discussed in Sec. III.
In Sec. IV the partition functions are obtained from the excitation energies,
and they are applied to the calculation of the abundances in free jet expansions 
of parahydrogen clusters.
A summary and some general conclusions are finally given in Sec. V.

\section{The diffusion Monte Carlo method}

The diffusion Monte Carlo (DMC) method~\cite{boronat02} solves the
imaginary-time Schr\"odinger equation for a function $f({\cal R},t) =
\Phi_{\rm var}({\cal R}) \Psi({\cal R}, t)$ which is the
product of the importance-sampling wave function $\Phi_{\rm var}$ and
the true ground-state wave function $\Psi$. The variable
${\cal R}$ represents the set $({\bf r}_1, \dots, {\bf r}_N)$
of $3N$ coordinates of the $N$
atoms which form the drop. The solution for $f$ is obtained by
constructing an approximate small-time Green's function $G({\cal
R},{\cal R}', \tau)$ which serves to advance $f$ by a small time step
$\tau$,
\begin{equation}
f({\cal R'},t+\tau) =\int d{\cal R} G({\cal R}',{\cal R}, \tau)f({\cal R},t).
\label{evolution}
\end{equation}
An initial set of walkers $\{{\cal R}_1, {\cal R}_2, \dots {\cal R}_{N_w}\}$
is created to represent the function $f({\cal R},0)$ at the initial time.
After many repeated applications of the short-time approximate Green's function,
all components of the starting wave function orthogonal to the ground state wave
function disappear and the contributions of the other eigenstates go 
exponentially to zero as the time goes to infinity. The remaining set of 
walkers then provides a valid representation of the converged solution.
In order to have a good statistical
accuracy it is convenient to continue the application of the Green's function
along a very large number of time steps and accumulate the resulting set of
walkers. Then the mixed estimators of physically interesting quantities,
such as the total energy, can be computed. Of course since the walkers are
strongly correlated a block average is carried out in order to estimate
the variance.

Our ground state calculations are based on a Jastrow-like importance sampling
wave function depending on just two parameters
\begin{equation}
\label{trialwf}
\Phi_{\rm var} = \prod_{i<j}^N \exp \left( - \frac {1}{2} \left( \frac{r_{ij}}{b}
\right)^5 - \frac{r_{ij}}{p} \right)~,
\end{equation}
where $r_{ij}=| {\mathbf r}_i-{\mathbf r}_j|$ is the distance between
the pair $(i,j)$.
Once the variational parameters were determined for each value of $N$,
DMC calculations were carried out with very small
real-time steps ($5\times 10^{-5}~{\rm K}^{-1}$ to $1\times 10^{-5}~
{\rm K}^{-1}$, for the light and heavier clusters, respectively) and very
long number of evaluation steps. Notice that the Bose symmetry
of the above guiding wave function is not modified in the DMC process.
The inclusion of triplet correlations in the importance sampling function
leads to a noticeable improvement of the variational energies, as showed
in Ref.~\cite{guardiola2008}. However, the DMC energies are essentially
the same as those obtained with only two-body correlations, with slightly
more binding in the heavier clusters. The main difference
lies in the reduction of the standard deviation by typically a factor of two.

Importance sampling wave functions with a non-zero angular momentum may equally
well be considered, and correspondingly the initial state will be an admixture
of eigenstates of a specified angular momentum $L$.
A convenient trial function with angular
momentum $L$ is the following
\begin{equation}
\label{trialwfL}
\Phi_{\rm var}^{(L)}({\cal R}) = \sum_{i=1}^{N}
{\rm Re} \left\{ |{\bf r}_i -{\bf R}_{\rm cm}|^L \, Y_{LL}(\bar{\omega}_i)
\right\} \times \Phi_{\rm var}({\cal R}) \, ,
\end{equation}
where $\bar{\omega}_i$ stands for the spherical angles of the vector
${\bf r}_i -{\bf R}_{\rm cm}$ which defines the position of the $i$-th particle
with respect to the center-of-mass coordinate
${\bf R}_{\rm cm}=\sum_i {\bf r}_i /N$, and $Y_{LM}$ is a spherical harmonic. 
Note that the sum over all the constituent particles and the subtraction of 
the center of mass coordinate ${\bf R}_{\rm cm}$ are required to maintain the 
the Bose symmetry and the translational invariance, respectively. Again, the 
application of the imaginary-time Green's function will select the lowest 
eigenstate in the subspace of angular momentum $L$, the remaining contributions 
being exponentially suppressed as the time increases. The prefactor entering 
Eq.~(\ref{trialwfL}) is identical to zero when $L=1$. In that
case we have used
\begin{equation}
\Phi_{\rm var}^{(1)}({\cal R}) = \sum_{i=1}^{N} {\rm Re} \left\{
|{\bf r}_i -{\bf R}_{\rm cm}|^3 \, Y_{11}(\bar{\omega}_i) \right\} \times
\Phi_{\rm var}({\cal R}) \, .
\end{equation}
A similar procedure to deal with $L$-states has been used in
Ref.~\cite{McMahon1993} for the $N=7$ cluster, and in Ref.~\cite{guardiola2006}
for $^4$He droplets.

Certainly, the trial function for $L \neq 0$ states is not positive definite and
consequently cannot be interpreted as a probability distribution function. We
have used the so-called fixed-node
approximation~\cite{reynolds82} assuming that the nodal surfaces are fixed by
the importance sampling function. In this approximation an exact energy for
the lowest angular momentum $L$ state is, in fact, not obtained. The
improved variational wave function and the corresponding mixed estimator
of the energy, which is derived, does, however, provide a variational
upper bound.

Besides the excited states with angular momentum $L$, one should also
consider vibrational excitations characterized by a radial quantum number $n$.
In short, let us represent each state by $(n,L)$. With the previous
trial wave function, the DMC method selects the lowest eigenstate $(0,L)$ in the
subspace of angular momentum $L$. An useful estimate of the energies of the
average vibrational excitations $(n \ge 1,L)$ in each $L$-subspace is obtained
by using sum rules~\cite{bohi79,lipp89}, which only require the knowledge of the
lowest $L$-state wave functions.
Consider the exact lowest state for a given angular momentum $L$ and the full
set of eigenstates of this subspace in order of increasing energy, represented
by $\Psi_0^{(L)}$ and $\{ \Psi_n^{(L)},E_n^{(L)}\}$, respectively, where
$n=0, 1 \dots$ is the vibrational quantum number. Let $Q({\cal R})$ be an
arbitrary operator, assumed to be scalar under rotations, for which
the sum rule of order $p$ is given by
\begin{equation}
\label{sum_rules}
M_p^{(L)} [Q] = \sum_{(n,\ell)\neq (0,L)} (E_n^{(\ell)} - E_0^{(L)})^p
 \left| \langle \Psi_n^{(\ell)} |Q| \Psi_0^{(L)} \rangle \right|^2 \, ,
\end{equation}
where the sum extends
over all eigenstates of the Hamiltonian with the exception of the lowest energy
state $(0,L)$ of angular momentum $L$.
An upper bound to the energy of the first excited state
of the subspace $L$ is obtained as
\begin{equation}
\label{upper_bound}
E_1^{(L)} - E_0^{(L)} \leq\frac{M_1^{(L)}[Q]}{M_0^{(L)}[Q]} \, ,
\end{equation}
which involves the energy weighted and the non-energy weighted sum rules.
Using the completeness relation of the set
of functions $\{ \Psi_n^{(L)}\}$, they can be written as
 \begin{eqnarray}
\label{eme0}
M_0^{(L)}[Q] &=& \langle\Psi_0^{(L)} |Q^2| \Psi_0^{(L)} \rangle
- \left| \langle \Psi_0^{(L)} |Q| \Psi_0^{(L)} \rangle \right|^2 \, , \\
\label{eme1}
M_1^{(L)} &=& \frac{\hbar^2}{2m}
\langle \Psi_0^{(L)} | \sum_{i=1}^N |
\nabla_i Q |^2 |\Psi_0^{(L)} \rangle \, .
\end{eqnarray}
The computation of these
expressions only requires a knowledge of the lowest state wave function of the
angular momentum $L$ subspace.

We present here results for the first
vibrational state of the subspace $L=0$, employing a general monopolar
operator
\begin{equation}
Q({\cal R}) = \sum_{p=1}^5 c_p \sum_{i < j} r_{ij}^p \, .
\end{equation}
The parameters $c_{p}$ have been optimized so as to obtain the
lowest upper bound to the first excited state. Notice that the excitations
obtained are approximate because
mixed matrix elements are used to calculate the required sum rules, so the
strict variational character is lost.

\section{The energy levels of parahydrogen clusters}
\label{sectionIII}

\begin{figure}[h!]
\includegraphics[width=0.7\linewidth,angle=-90]{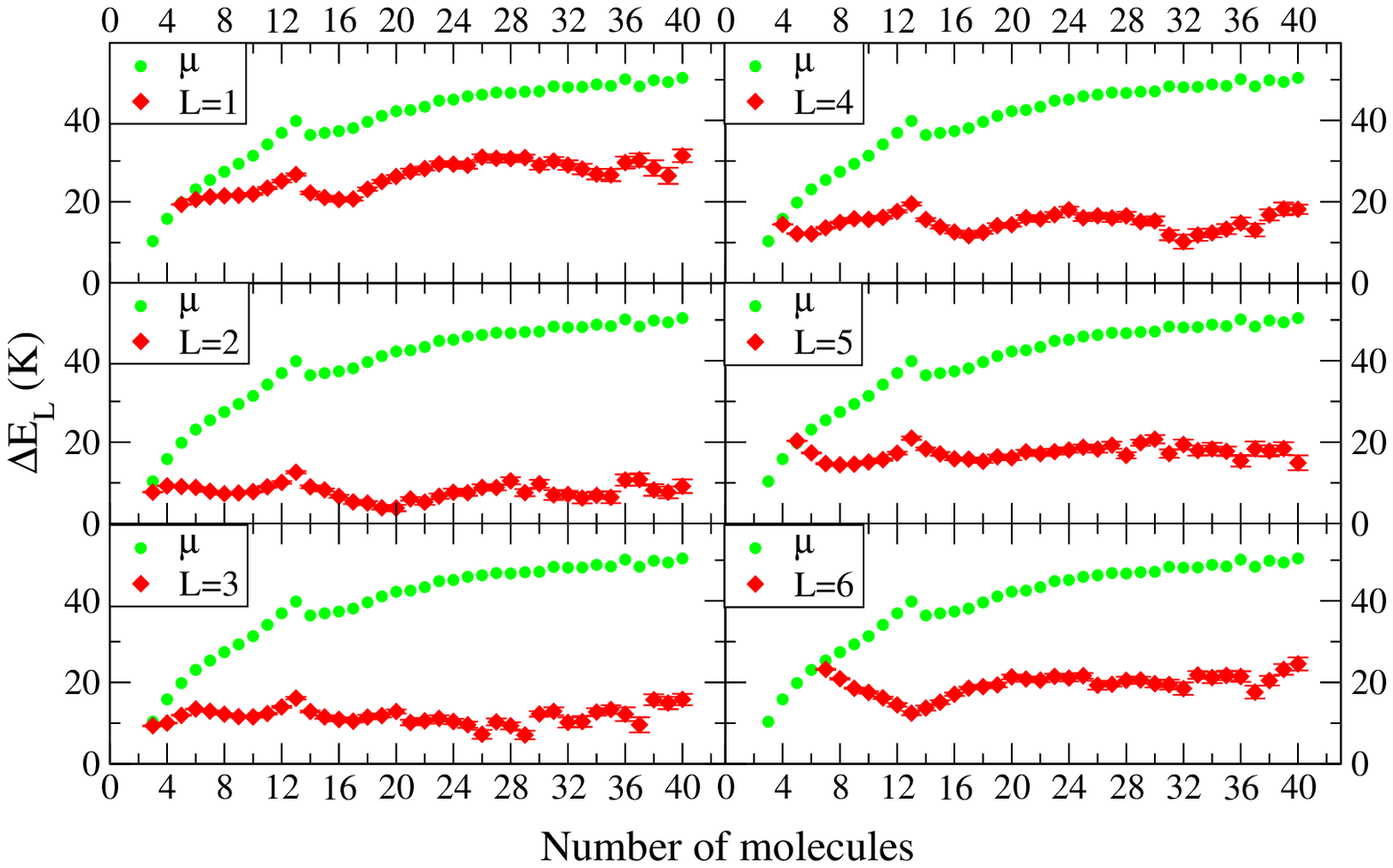}
\caption{(Color online)
Excitation energies (in K) of (pH$_2$)$_N$ clusters as a function of
the number $N$ of constituents for states with $L=1-6$. The
chemical potential $\mu(N)$ is also displayed to indicate the stability
limit.}
\label{fig1}
\end{figure}

\begin{figure}[h!]
\includegraphics[width=0.7\linewidth,angle=-90]{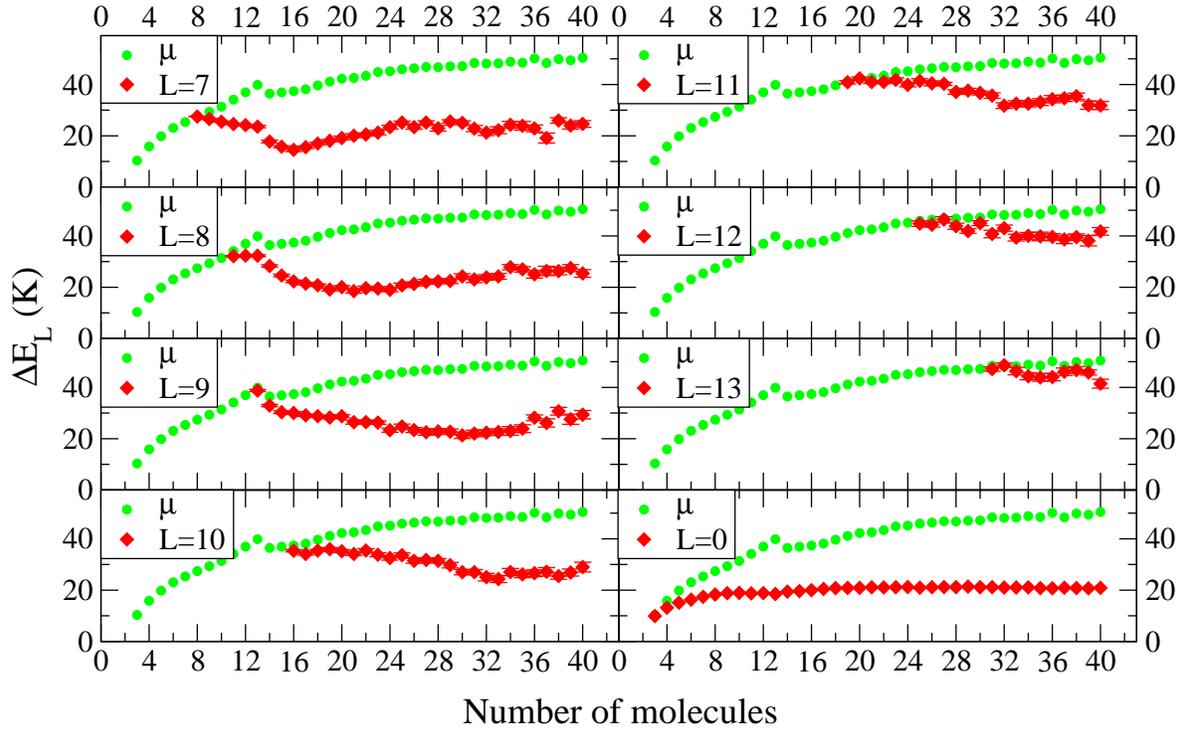}
\caption{(Color online)
Excitation energies (in K) of (pH$_2$)$_N$ clusters as a function of
the number $N$ of constituents for states with $L=7-13$ and the vibrational
state with $L=0$. The chemical potential $\mu(N)$ (green dots) is also
displayed to indicate the stability limit.}
\label{fig2}
\end{figure}

\begin{figure}[h!]
\includegraphics[width=0.7\linewidth,angle=-90]{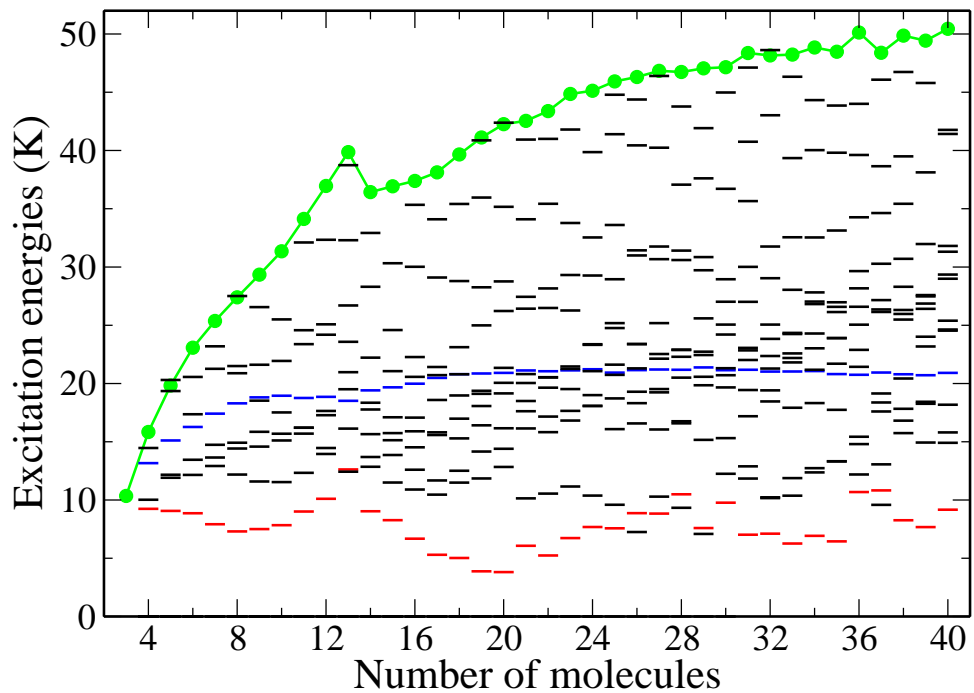}
\caption{(Color online)
Excitation energies (in K) of (pH$_2$)$_N$ clusters as a function of
the number of constituents. The chemical potential $\mu(N)$ (green dots) is also
displayed to indicate the stability limit. Levels with $L=2$ and 0 are plotted in
red and blue, respectively.}
\label{fig3}
\end{figure}

Once calculated for each cluster the ground state $E_0^{(0)}(N)$ and the excited
state energies $E_0^{(L)}(N)$ and $E_1^{(0)}(N)$, the excitation energies
are defined as
\begin{eqnarray}
\label{notation1}
\Delta E_{L\neq0}(N) &=& E_0^{(L)}(N) - E_0^{(0)}(N) \, , \\
\label{notation2}
\Delta E_0(N) &=& E_1^{(0)}(N) - E_0^{(0)}(N) \, .
\end{eqnarray}
They are plotted as a function of $N$ in figures~\ref{fig1}-\ref{fig3}.
For the sake of clarity, we have displayed them separately for each
value of $L$ in figures~\ref{fig1}-\ref{fig2} and the full
spectra is showed in figure~\ref{fig3}. The chemical potential or dissociation energy
\begin{equation}
\mu(N) = E_0^{(0)}(N-1)-E_0^{(0)}(N) \, ,
\end{equation}
defined as the difference between the ground state energies of neighbor
clusters, has also been plotted in all cases. Indeed the chemical potential
defines the energetic region in which excited states are stable. Only stable states
have been plotted in figures~\ref{fig1}-\ref{fig3}.

As previously discussed in Refs.~\cite{guardiola2006a,guardiola2008}, the most
prominent result regarding the chemical potential is the enhanced stability
of the cluster $N=13$, which confirms its magical character.
A mild peak appears also at $N=36$, but one cannot exclude that it could
simply be a statistical fluctuation. An additional confirmation of its
magicity will be given later on. 
In contrast, PIMC calculations of Refs.~\cite{mezzacapo2007} and~\cite{khairallah2007},
at T=1 and 0.5~K respectively, show an enhanced stability at $N=13$, 26, 29,
34, and 39.

The main characteristics of the spectra displayed in figures~\ref{fig1} and \ref{fig2} 
can be summarized as follows. All clusters exhibit stable 
excitations for $L=2$, 3 and 0 at $N=3$ and onwards. The next excited level 
with $L=4$ starts to be bound at $N=4$, the following levels with $L=5$ and 
1 appear at $N=6$. The next $L=6, \dots$ levels appear at regularly
increasing size thresholds. In the considered size range the highest stable 
excited level corresponds to $L=13$, which appears at $N=31$ and onwards.
The quadrupolar $L=2$ excitation is the lowest one for all sizes, except at
$N=26$, 28, 29 and 37, for which the octupolar $L=3$ state lies at a slightly
lower energy.
The excitations with $L \ge7$ do not cross each other: the higher the angular
momentum the higher the excitation energy.
The vibrational excitation energies of the state $(n=1,L=0)$ present a smooth 
behavior as a function of N, apart from a mild minimum at $N=13$. We have not 
pushed our calculations to the vibrational states $(1,L \neq 0)$. Presumably, 
they would also present a monotonic behavior with respect to their 
corresponding $(0,L)$ states.
In contrast, the excitation energies of the states $(n=0,L \neq 0)$
have a non-monotonic behavior with $N$. 

The $N=13$ cluster is very peculiar. Its excitation energies present a prominent 
peak for the states with $L=1-5$,
a prominent minimum for $L=6$, and a pronounced change of trend for $L=7$, 
8 and 9. This cluster has no stable excited states for $L \ge 9$.
No other so enhanced bumps appear up to $N=40$.
Such a behavior is related to the truly magic 
character of this size. Its enhanced stability results in a repulsion of 
the excitations, similarly to what has been observed in classical Lennard-Jones
calculations~\cite{honeycutt1987,frantz2001}.
The prominent minimum found for $L=6$ at $N=13$ means that this value of 
the angular momentum is also special. A similar phenomenon was observed
in the study of doped parahydrogen clusters CO@{(pH$_2$)}$_N$ by reptation 
QMC calculations~\cite{baroni2005}, where the time-correlation functions of 
the various multipole moments in the rotating-axes frame have been computed. 
A persistent crystal structure for $N=12$ and a persistent time correlation 
of the $L=6$ multipole has been observed by these authors. Their explanation 
for these facts was based in the property that the only non-null multipolar 
moments for the $N=13$ cluster with icosahedral symmetry are $L=6$ and $L=10$.

Our results are compared in Table~\ref{table1} with those of
McMahon {\it et al.}~\cite{McMahon1993} for $N=7$ which, to the best of our
knowledge, are the only available calculations of excited states of parahydrogen
clusters. These authors have performed a DMC calculation using the BHKOS
interaction and a trial wave function for $L$-excited states analogous
to our Eq.~(\ref{trialwfL}). It can be seen in the table that both 
calculations are in substantial agreement within at most two standard 
deviations.

In figure~\ref{fig3} are collected our calculated DMC excitation energies, 
omitting error bars for the sake of clarity. The obtained 
spectrum for each (pH$_2$)$_N$ cluster is {\em complex}, in the
sense that it does not resemble any of the simple models of excitations.
The states with $L=2$ are the lowest lying, apart from the few 
cases mentioned before. It is worth mentioning the presence of relevant 
energy gaps at $N=13$ and $N=36$, in coincidence with peaks in the chemical 
potential. Both facts indicate a magical character of these
clusters. Note that analogous energy gaps have been determined in
classical Lennard-Jones clusters~\cite{honeycutt1987}.

There is almost no information, neither experimental nor theoretical, of 
the spectrum of clusters weakly bound by van der Waals like interactions.
An indirect determination of the (pH$_2$)$_2$ dimer excitations has been 
reported in Refs.~\cite{watanabe1964,mckellar90, mckellar1991}. These 
interesting experiments refer to the absorption of infrared radiation by 
parahydrogen gas at 20~K, slightly above the triple point. Just focusing 
in the region around the $Q_1(0)$ line there has been  observed the 
{\em absence} of absorption at exactly the frequency of the free $Q_1(0)$ 
line, and the presence of two satellites at energies above and below the 
free parahydrogen frequency, interpreted as a signal of the dimer excited 
states. In practice, absorption experiments provide an indirect way
of detecting the two-body ground state pH$_2-$pH$_2$ and measuring its 
spectrum. 

In the case of $^4$He droplets, an indirect determination of their 
excitation energies results from the temperature-dependent analysis of 
the production abundances of such clusters in ultrasonic expansion of 
pressurized gas~\cite{guardiola2006}. Sudden jumps have been observed in 
the quantum partition function at cluster sizes which can just
accommodate one more additional stable excitation. The nice agreement 
between the experimental magic sizes and theoretical stability thresholds 
provides a confirmation of the calculated energy levels of these clusters.

In principle one could expect some parallelism between the spectra of 
$^4$He droplets and those made of parahydrogen: both constituents are bosons 
and the interaction is rather similar. However, our calculations rule out 
such an analogy, as helium clusters present regularly spaced spectra.
In contrast, no simple pattern can be guessed for the spectra of 
parahydrogen clusters. 
Only the excitation energy of the $L=0$ level has a smooth behavior with 
the number of pH$_2$ molecules, with a small inflection at $N=13$. Its 
value is roughly constant near 20~K, quite far away from the dissociation 
limit, in contrast with the analogous level in $^4$He droplets which is 
close to the dissociation limit.

The $(n=0,L>1)$ of $^4$He excitations correspond to surface modes, described 
by ripplon like oscillations with radial frequencies
given by~\cite{rayleigh1879, bohr1998,stringari1990}
\begin{equation}
\omega= \left[ \frac{\sigma_0}{m\rho_0R_0^3} L(L-1)(L+2)\right]^{1/2} \, .
\end{equation}
These frequencies were corrected by Tamura~\cite{tamura1997} to include 
finite size effects and the corrected expression reproduces very well the 
calculated DMC spectra of helium clusters~\cite{guardiola2006}.
Certainly, the smooth behavior related to the angular momentum dependence
$[L(L-1)(L+1)]^{1/2}$ does not emerge in our calculated spectra of pH$_2$ 
clusters.

The excitation energy of the translationally invariant {\em forbidden} 
$1$ mode, which requires a nodal surface in the radial coordinate space, 
has a quite high energy compared with the {\em allowed} levels  
($2$, $3$, \dots), like in the case of helium droplets. Nevertheless, its 
energy is again far away from the dissociation limit, in contrast with the 
case of $^4$He clusters. With respect to the variation with the number of
molecules, it is noticeably smoother than the levels of low angular momentum 
(up to $L=4$) but reveals a prominent bump near $N=13$.

At the light of our calculations we infer a coexistence of liquid and 
crystal structure near $N=13$, the liquid part being the responsable of the 
$L=2, 3, ...$ excitations, and the crystal part the responsable of the 
$L=6$ excitation. The region around $N=13$ where the energy of the $L=6$ 
mode is decreasing up to $N=13$ and afterwards increasing up to a plateau 
will correspond to partial icosahedral symmetry. We have analyzed the known 
crystal structures related to Lennard-Jones clusters reported in the 
Cambridge Cluster Data Base~\cite{CCDB}, by computing the multipolar moments. 
There are three significative cases worth to be mentioned.
Firstly, the $N=19$ cluster, for which the multipolar moments
$L=3$ and $L=4$ are null and for which we obtain a very low $L=2$ excitation 
and much higher $L=3$ (a peak) and $L=4$. Secondly, the $N=26$ case, whose 
first non-null multipole is $L=3$ and appears in our calculations with a 
minimum for this excitation. Finally, the $N=38$ cluster, whose first 
non-null multipole is $L=4$ appearing again in our calculations a valley 
around this value of the number of constituents. In conclusion, even if we 
do not find a simple or familiar model to describe the excitations of 
parahydrogen clusters, some aspects of the spectrum are qualitatively described 
by assuming a coexistence of liquid and solid phases.

\section{Thermal effects and magic numbers}

The knowledge of the full spectra allows us to analyze different thermal 
effects by means of the partition function. In particular, we consider in 
this section the expectation value of the temperature-dependent energy 
$E(T)$ and the effect of excited states on the production abundances of 
clusters in free jet expansion of pressurized gas.

The partition function of each cluster is expressed as the product
\begin{equation}
\Xi_N = Z_N \xi_N
\end{equation}
of the center-of-mass ($Z_N$) and internal ($\xi_N$) partition functions.
The former can be written as
\begin{equation}
Z_N = \left( \frac{NmkT}{2 \pi \hbar^2} \right)^{3/2} \, .
\end{equation}
The internal partition function is given in terms of bound state energy levels
$E_j$ of the state $j=(n,L)$ of the N-molecule cluster by
\begin{equation}
\xi_N = \sum_j g_j(N) \exp \left[-\frac{E_j(N)}{kT} \right] \, ,
\end{equation}
where the degeneracy factor for the state with the radial quantum
number $n$ and angular momentum $L$ is given by $g_j=2L+1$.

By using the notation of section \ref{sectionIII}, 
Eqs.~(\ref{notation1}-\ref{notation2}),   
the internal partition function may be written in a computationally 
convenient form
\begin{equation}
\xi_N = \exp\left[-\frac{E^{(0)}_0(N)}{kT} \right]
\left[ 1 + \sum_L (2L+1)  \exp \left[-\frac{\Delta E_L(N)}{kT} \right]\right] \, ,
\end{equation}
where $\Delta E_L(N)$ are the excitation energies for angular momentum $L$, 
and the statistical factor has been written in the explicit form $(2L+1)$.

The internal energy of a given cluster at a temperature $T$ is given by
\begin{equation}
E_N(T) = E^{(0)}_0(N) +
\frac
{ 1 + \sum_L (2L+1) \Delta E_L(N) e^{ -{\Delta E_L(N)}/{kT}}}
{ 1 + \sum_L (2L+1)  e^{ -{\Delta E_L(N)}/{kT}}} \, .
\end{equation}
Sizable thermal effects on the energy values will start to appear at temperatures 
close to the energy of the first excited state. Thus, at $T=1$ the energy 
changes are minimal, with a maximum of 0.4~K for $N \simeq 20$, related to 
the minimum of $\Delta E_2$ (see fig.~\ref{fig1}). The energy of the other 
clusters remains practically unaltered, and consequently the
chemical potential at $T=1$ has practically the same shape as at $T=0$. 
This observation is interesting
in relation with PIMC calculations, which have been carried out
at $T=0.5$~K~\cite{khairallah2007} and $T=1$~K~\cite{mezzacapo2007}.
The series of peaks observed in the PIMC chemical potential as a function 
of the number of molecules is not reproduced by the present extension of DMC 
to $T=1$. 
Although different interactions have been used in these calculations, namely
BHKOS in DMC and SG in PIMC, we do not expect this as the origin of
the differences. Indeed, DMC results using both interactions~\cite{guardiola2008}
give substantially the same chemical potential shape, apart from the mentioned
mild peak at $N=36$, not visible with the SG interaction. 

On the other hand, the effect of the temperature on the mechanism by 
which clusters are formed in the free jet expansion turns out to be very interesting. 
The analysis which follows is similar to the one carried out for Helium 
droplets~\cite{bruhl2004,guardiola2006} which explained the experimentally 
observed anomalously high production rates of clusters with a specific
number of constituents, the anomalies being related to the excitation 
spectrum of the clusters and not to a magical character.

Let us assume that cluster growth is dominated by the chemical 
equilibrium reaction
\begin{equation}
(pH_2)_{N-1}+pH_2+X \rightleftharpoons  (pH_2)_N+X \, ,
\end{equation}
where $X$ is a spectator particle required to fulfill the energy and momentum
conservation laws. The related equilibrium constant is independent of the 
spectator particle and is given by a quotient of partition functions
\begin{equation}
K_N = \frac{\Xi_N}{\Xi_{N-1} \Xi_1} \, .
\end{equation}

\begin{figure}[h!]
\includegraphics[width=0.7\linewidth,angle=-90]{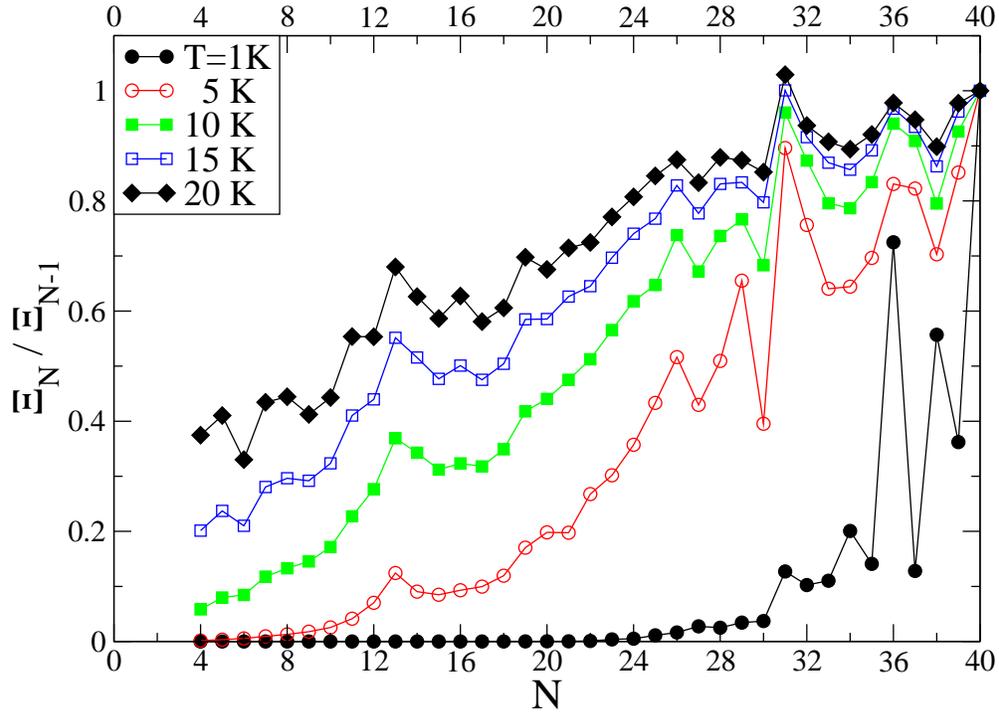}
\caption{(Color online)
The ratio of the partition functions $\Xi_N/\Xi_{N-1}$ is plotted versus $N$
for several values of temperature: 1, 5, 10, 15, and 20~K. The scale has been
arbitrarily fixed at $N=40$.}
\label{fig4}
\end{figure}

The equilibrium constant is dominated by the quotient $\Xi_N/\Xi_{N-1}$, 
which is conveniently written as
\begin{equation}
\frac{\Xi_N}{\Xi_{N-1}} = \left(\frac{N}{N-1}\right)^{3/2}
e^{\mu(N)/kT} \,
\frac{\displaystyle 1+\sum_L g_L e^{-\Delta E_L(N)/kT}}
{\displaystyle 1+\sum_L g_L e^{-\Delta E_L(N-1)/kT}}
\end{equation}
Neither the chemical potential nor the excitation energies are smooth functions
of $N$, as has been show in figure~\ref{fig3}. One can expect that 
the above quotient will reflect such a non monotonic behavior as jumps at particular 
values of $N$. The quotient of partition functions $\Xi_N/\Xi_{N-1}$ 
is displayed in figure~\ref{fig4} as a function of $N$ for several values of 
temperature, lower than the source temperatures of Ref.~\cite{tejeda2004},
about 40~K. 
As these ratios span several orders of magnitude as $T$ varies, we have arbitrarily 
given the value 1 to the case $N=40$.

Clearly pronounced peaks are observed at $N=13$, 31  and $36$, and less 
pronounced at $N=26$. Other peaks exist, like $N=19$ and $N=29$, but they
only appear at specific values of $T$.
The pronounced peaks give an explanation of the experimentally determined 
production peaks at $N=13$ and 33. It should be mentioned that
the relation between the Raman shift and the number of molecules of the 
clusters in the work of Ref.~\cite{tejeda2004} is an 
{\em extrapolation} from small values of $N$, and the position of the peak 
at $N = 33$ may be imprecise. It is worth noticing that the last value is
half way between the theoretical peaks at $N=31$ and $N=36$ related 
to an enhanced production.

\section{Conclusions}

In this work we have determined the excited-state energy levels of parahydrogen 
clusters containing up to 40 molecules by means of a DMC calculation based
on an importance-sampling function with a nonzero angular momentum $L$ plus
the fixed-node approximation. A variational upper bound has been obtained for
the lowest-lying energy of a given $L$-state. Bound excited states have been 
found for states $(n=0,L=1-13)$ and $(n=1,L=0)$, the size threshold for binding
increasing as $L$ increases. It turns out that the spectra are  
{\em complex}, in the sense that no simple pattern can be guessed to fit them 
as a function of the angular momentum of the state and the size number. 
In particular, and contrary to what happens in the analogous 
system of helium droplets, the liquid drop model formula is excluded.  

The $N=13$ and $N=36$ clusters are magic, as indicated by the existence of peaks
in the chemical potential in coincidence with energy gaps for their first
low-lying excited state, relative to their neighbors.

The spectra of weakly bound van der Waals clusters have not yet been 
experimentally measured. Only an indirect 
determination~\cite{bruhl2004,guardiola2006} has resulted from the analysis 
of enhanced production of clusters of $^4$He in terms of the successive 
stabilization of excited levels at growing values of the number of 
constituents of the clusters.
The analysis of the partition function of parahydrogen clusters has revealed
the existence of production enhancements at $N=13$, 31 and 36, in nice 
agreement with the experiment~\cite{tejeda2004}. The effect is similar to
the case of helium clusters, but the nature is very different. Instead of the
stability threshold reason of helium clusters, here the enhanced
production is basically related to the existence of energy gaps in the 
excitations at the mentioned values of the number of constituents.
There may be some corrections due to not including the $(n=1,L \ne 0)$ levels 
in the partition function. We expect, however, 
these effects to be small, as presumably such vibrational excitations would 
be some 20~K above the non-vibrational $(n=0,L)$ corresponding level.

\begin{acknowledgments}
Stimulating conversations  with S. Montero, M. Sevryuk and J.P. Toennies
are gratefully acknowledged.
This work has been supported by MCyT (Spain), grant No. FIS2007-60133.
\end{acknowledgments}

\hfill
\newpage

\begin{table}[h!]
\caption{\label{table1}
Cluster with $N=7$. DMC results of Ref.~\cite{McMahon1993} are compared
with present results. All energies are in K}
\begin{ruledtabular}
\begin{tabular}{ccc}
        & McMahon {\it et al.} &  Present results   \\ \hline
$E_0^{(0)}$ & -98.89(15) & -98.76(1) \\
$\mu$       & 25.90(15)  & 25.37(2)  \\
$\Delta E_2$ & 8.3(3)    &  7.92(13) \\
$\Delta E_3$ & 13.2(2)    &  12.92(17) \\
$\Delta E_5$ & 15.4(2)    &  14.74(18) \\
\end{tabular}
\end{ruledtabular}
\end{table}

\hfill
\newpage

\end{document}